\documentclass[aps, prl,
reprint,
 amsmath,amssymb,
superscriptaddress,
twocolumn
]{revtex4-2}

\usepackage{gensymb}
\usepackage[caption=false]{subfig}

\usepackage{graphicx}
\usepackage{dcolumn}
\usepackage{bm}
\usepackage{float}
\usepackage{hyperref}
\usepackage{cleveref}
\usepackage{enumerate}
\usepackage{dsfont}
\usepackage{natbib}
\usepackage[T1]{fontenc}
\usepackage[polish, english]{babel}
\usepackage[utf8]{inputenc}
\usepackage{array}
\usepackage{booktabs}
\usepackage{diagbox}
\usepackage{multirow}
\usepackage[table]{xcolor} 

\definecolor{darkblue}{RGB}{0, 0, 140}
\definecolor{tab:pink}{cmyk}{0.1156,0.6411,0,0}
\definecolor{tab:blue}{cmyk}{0.8481, 0.4814, 0.0596, 0.0006}
\definecolor{tab:bluelight}{cmyk}{0.2040, 0.0679, 0.0119, 0}
\definecolor{tab:pinklight}{cmyk}{0.0173, 0.2949, 0, 0}

\newcommand{\ch}[1]{\textcolor{black}{#1}}

\begin{document}

\title{Machine Learning of Implicit Combinatorial Rules in Mechanical Metamaterials}

\author{Ryan van Mastrigt}
\email{r.vanmastrigt@uva.nl}
\affiliation{Institute of Physics, Universiteit van Amsterdam, Science Park 904, 1098 XH Amsterdam, The Netherlands}
\affiliation{AMOLF, Science Park 104, 1098 XG Amsterdam, The Netherlands}

\author{Marjolein Dijkstra}

\affiliation{Soft Condensed Matter, Debye Institute for Nanomaterials Science, Department of Physics, Utrecht University,
Princetonplein 5, 3584 CC Utrecht, The Netherlands}

\author{Martin van Hecke}

\affiliation{AMOLF, Science Park 104, 1098 XG Amsterdam, The Netherlands}
\affiliation{Huygens-Kamerling Onnes Lab, Universiteit Leiden, Postbus 9504, 2300 RA Leiden, The Netherlands}

\author{Corentin Coulais}%
\affiliation{Institute of Physics, Universiteit van Amsterdam, Science Park 904, 1098 XH Amsterdam, The Netherlands}

\date{\today}
\begin{abstract}{
Combinatorial problems arising in puzzles, origami, and (meta)material design have rare sets of solutions, which define complex and sharply delineated boundaries in configuration space. These boundaries are difficult to capture with conventional statistical and numerical methods.
Here we show that convolutional neural networks can learn to recognize these boundaries for combinatorial mechanical metamaterials, down to finest detail, despite using heavily undersampled
training sets, and can successfully generalize.
This suggests that the network infers the underlying combinatorial rules from the sparse training set, opening up new possibilities for complex design of (meta)materials.
}
\end{abstract}

\maketitle

From proteins and magnets to metamaterials, all around us systems with emergent properties are made from collections of interacting building blocks. Classifying such systems---do they fold, are they magnetized, do they have a target property---normally involves calculating these properties from their structure. This is often straightforward in principle, yet computationally expensive in practice, \textit{e.g.} requiring the diagonalization of large matrices. Machine learning algorithms such as neural networks (NNs) forgo the need for such calculations by ``learning'' the classification of structures. In particular, machine learning has proven successful to find patterns in crumpling~\cite{hoffmann2019machine}, active matter~\cite{Colene2016708118, falk2021learning, dulaney2021machine} and hydrodynamics~\cite{bar2019learning}, \ch{photonics}~\cite{wiecha2021deep, ma2021deep, xu2021interfacing}, predict structural defects and plasticity~\cite{harrington2019machine, mozaffar2019deep}, design metamaterials~\cite{bessa2019bayesian, gu2018novo, wang2020deep, coli2022inverse, Basteke2111505119, shin2021spiderweb, hanakata2020forward, forte2022inverse}, determine order parameters~\cite{cubuk2015identifying, bapst2020unveiling, schoenholz2016structural, hsu2018machine, venderley2018machine, swanson2020deep, van2020classifying, miles2021correlator}, identify phase transitions~\cite{andrejevic2020machine, carrasquilla2017machine, van2017learning, deng2017machine, zhang2017quantum, zhang2018machine, zhang2019machine, ch2017machine, rem2019identifying, bohrdt2019classifying, carrasquilla2020machine, van2018learning, bohrdt2021analyzing, sigaki2020learning,geiger2013neural, dietz2017machine, zhang2021machine, coli2021artificial}
, and predict protein structure~\cite{jumper2021highly}.
In these examples, the relevant property typically varies smoothly and there is no sharp boundary separating classes in configuration space. NNs are thought to be successful because they interpolate these blurred boundaries, even when the configuration space is heavily undersampled.
\begin{figure}[b!]
    \centering
    \includegraphics{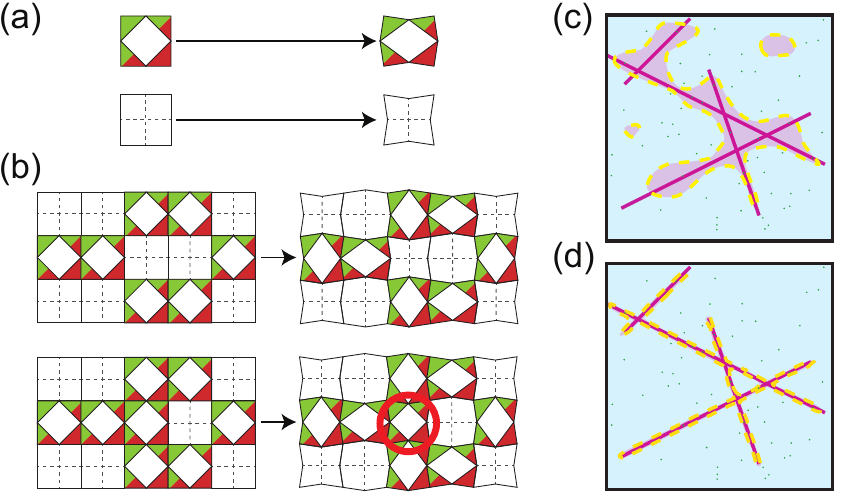}
    \vspace{0.01eM}
    \caption{(color online)
    (a) The building block of \cite{coulais2016combinatorial} can be tiled in two orientations (left) that have a distinct deformation in two dimensions (right). 
    (b) The building blocks of (a) combine into larger designs (structures) that are either C (top) or I (bottom). A change of a single building block can frustrate the deformation (red circle) and change the structure from one that hosts a zero mode (a deformation that costs no energy) (C) to one that does not host a zero mode (I).
    A set of rules can be formulated for a unit cell design to have a zero mode~\cite{coulais2016combinatorial}.
    (c, d) Conceptual configuration spaces of a discrete combinatorial metamaterial problem. Class C (pink lines) exists in a background of class I (blue), is sensitive to perturbations, and has a complex filamentous structure. 
    Distinguishing between a network with a ``coarse'' decision boundary (purple dashed line) (c) versus a network with a ``fine'' decision boundary (d) is not possible with the test set (green dots) due to the undersampled C-I boundary.}
    \label{fig:smiley_designspace}
\end{figure}

In contrast, combinatorial problems, \textit{viz.} those where building blocks have to fit together as in a jigsaw puzzle, feature a sharp boundary between compatible (C) and incompatible (I) configurations. 
Such problems arise in self-assembly~\cite{gazit2007self, levin2020biomimetic}, folding~\cite{hull2002combinatorics, dieleman2020jigsaw}, tiling problems~\cite{demaine2007jigsaw} and combinatorial mechanical metamaterials~\cite{coulais2016combinatorial, meeussen2020topological, coulais2018multi, bossart2021oligomodal}. The latter are created by tiling different unit cells and are restricted by kinematic compatibility. A simple example is that of structures that can be either floppy (zero mode) or frustrated (no zero mode) (Fig.~\ref{fig:smiley_designspace}(a, b)). The floppy structures require a specific arrangement of building blocks where all the deformations fit together compatibly (C), and therefore are rare and very sensitive to small perturbations. These perturbations are likely to induce frustrated incompatible (I) configurations (Fig.~\ref{fig:smiley_designspace}(b)). The space of C designs can be pictured as needles in a haystack (Fig.~\ref{fig:smiley_designspace}(c, d)) and crucially is determined by a set of implicit combinatorial rules. Unless we know these rules, these problems are typically computationally intractable.

Here we show that convolutional neural networks (CNNs) are able to accurately perform three distinct classifications of combinatorial mechanical metamaterials and to generalize to never-before-seen configurations. Crucially, we find that well-trained CNNs can capture the fine structure of the boundary of C, despite being trained on sparse datasets.
These results suggest that CNNs implicitly learn the underlying rule-based structure of combinatorial problems.
This opens up the possibility for using NNs for efficient exploration of the design space and inverse design when the combinatorial rules are unknown.

\emph{Coarse vs. fine boundaries---} 
The boundary between C and I configurations has the shape of needles in a haystack. Therefore, in a randomly sampled training set, this boundary will be typically undersampled, \emph{e.g.} the training set will contain few I close to C (see SM~\footnote{See Supplemental Material at [URL will be inserted by publisher] for more details on the undersampled C-I boundary in the training sets.\label{fn: undersample C-I}}). We argue that a NN simply interpolating the training data will misclassify most I configurations close to C, resulting in a ``coarse'' decision boundary around C (Fig.~\ref{fig:smiley_designspace}(c)). 
Instead, an ideal NN should approximate the fine structure of the needles more closely, resulting in a ``fine'' decision boundary around C (Fig.~\ref{fig:smiley_designspace}(d)).
While this may sound impossible, 
let's recall that this fine structure ultimately arises from combinatorial rules. 
These rules are in principle much simpler than the myriad of compatible configurations C they can generate.
Hence,
the question is whether NNs could implicitly learn these rules and finely approximate the fine boundary with great precision.
Although a NN can classify perfectly the metamaterial M1 of Fig.~\ref{fig:smiley_designspace}(a, b) (Tab.~\ref{tab:confusion matrices}), this is not sufficient to address this question because the data set is too small and the C configurations are too rare to consider larger configurations (see SM~\footnote{See Supplemental Material at [URL will be inserted by publisher] for more details on the design rules and rarity of the metamaterial in Fig.~\ref{fig:smiley_designspace}.}).

\begin{table}[b!]
    \centering
    \caption{Confusion matrices of trained CNNs with the lowest validation loss over the test set for the classification problems of Fig.~\ref{fig:smiley_designspace}(b) (M1), Fig.~\ref{fig:modescaling}(d) (M2.i), and Fig.~\ref{fig:modescaling}(e) (M2.ii).}
    \label{tab:confusion matrices}
    \begin{tabular}{*{10}{c}}
    && \multicolumn{2}{c}{\textbf{M1}}& \hspace{0.1cm} & \multicolumn{2}{c}{\textbf{M2.i}} & \hspace{0.1cm} & \multicolumn{2}{c}{\textbf{M2.ii}} \\
      && \multicolumn{2}{c}{predicted} && \multicolumn{2}{c}{predicted} && \multicolumn{2}{c}{predicted} \\
       \multicolumn{2}{c}{}& \textcolor{tab:pink}{C} & \textcolor{tab:blue}{I} && \textcolor{tab:pink}{C} & \textcolor{tab:blue}{I} && \textcolor{tab:pink}{C} & \textcolor{tab:blue}{I} \\
      \multirow{2}{*}{actual} & \textcolor{tab:pink}{C} & \cellcolor{tab:pink} 19 & \cellcolor{tab:pinklight} 0&& \cellcolor{tab:pink} 685 & \cellcolor{tab:pinklight} 1 && \cellcolor{tab:pink}43418 & \cellcolor{tab:pinklight} 750\\
      & \textcolor{tab:blue}{I} & \cellcolor{tab:bluelight} 0 & \cellcolor{tab:blue}4896 && \cellcolor{tab:bluelight} 29 & \cellcolor{tab:blue} 149265 && \cellcolor{tab:bluelight} 453 & \cellcolor{tab:blue} 105361\\
    \end{tabular}
\end{table}

\begin{figure}[b!]
\centering
\includegraphics[width=\columnwidth]{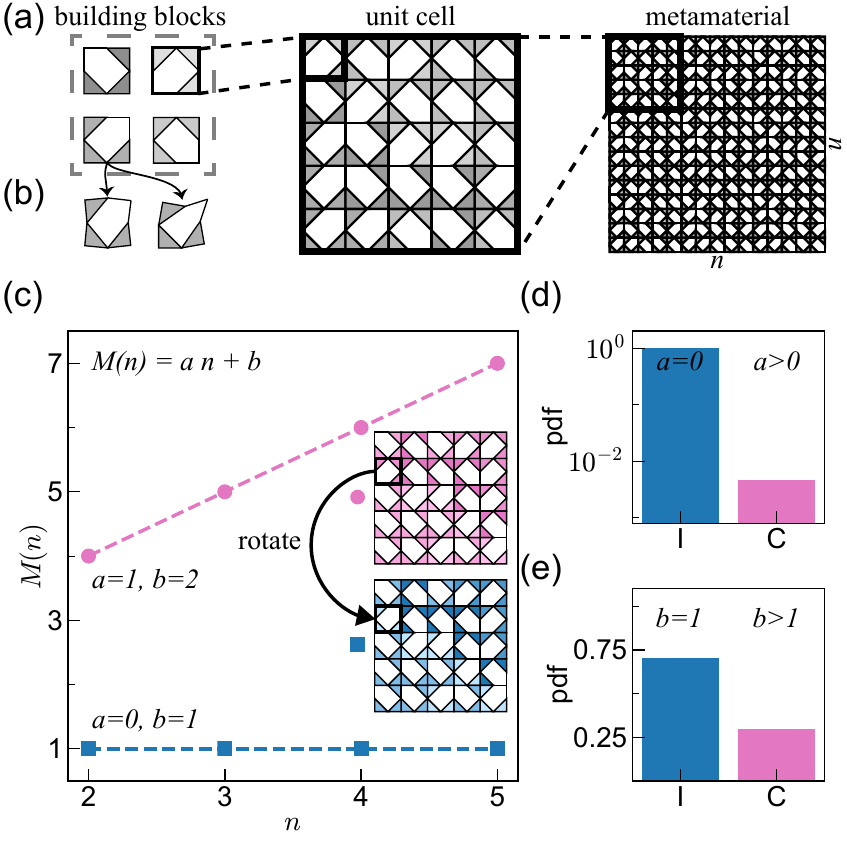}
\caption{
(color online)
(a) Four two-dimensional
building blocks (left), combined into a square $5\times 5$ unit cell (middle), which is tiled on a $n=3$ grid, form a combinatorial metamaterial (right). (b) The building blocks feature two zero modes and four orientations with distinct deformations.
(c) The number of zero modes $M(n)$ as function of $n$ for two unit cells. The pink unit cell (circles) differs by a point mutation from the blue unit cell (squares), yet the pink unit cell has $a=1$ and $b=2$ and the blue unit cell has $a=0$ and $b=1$. Thus the pink unit cell is classified as class C for both classification problems while the blue unit cell is classified as class I for both problems.
(d) Probability density function (pdf) for classification problem (ii). Class C is more rare than class I.
(e) Probability density function (pdf) for classification problem (i). Class C is much rarer than class I.}
\label{fig:modescaling}
\end{figure}

\textit{Metamaterial Classification}\textemdash Therefore, to see if NNs are still able to learn the structure of C if the C-I boundary is undersampled, we consider another combinatorial metamaterial M2 ~\cite{bossart2021oligomodal}
(for details on how we define it, see Fig.~\ref{fig:modescaling}(a, b)).
While metamaterial M1 had a unit cell of size $k\times k$ with $k=1$, metamaterial M2 has larger unit cell size---we focus on $k=5$ in the Main Text and cover the cases $k=3$ to $8$ in the SM. For such a metamaterial, the design space is too large to fully map and class C is rare, yet class C is abundant enough that we can create sufficiently large training sets to train NNs. 

The number of zero modes $M(n)$ of a metamaterial consisting of $n \times n$ unit cells depends on the design of the unit cell: when the linear size $n$ is increased, the number of zero modes $M(n)$ either grows linearly with $n$ or saturates at a non-zero value (Fig.~\ref{fig:modescaling}(c)) as $M(n)=a n + b$, where $a$ and $b$ are positive integers.
Accordingly, we can now specify two well-defined binary classification problems, which each feature a rare ``compatible'' (C) class and frequent ``incompatible'' (I) class (Fig.~\ref{fig:modescaling}(d, e)): (i) $a>0$ (C) vs. $a=0$ (I). The metamaterial with $a>0$ hosts zero modes that are organized along strips, for which one can formulate combinatorials rules (see SM~\footnote{See Supplemental Material at [URL will be inserted by publisher] for a detailed description of and numerical evidence for combinatorial rules of classification (i).}); (ii)  $b>1$ (C) vs. $b=1$ (I). The metamaterial with $b>1$ hosts additional zero modes\ch{---up to 6---}that typically span the full structure and for which the rules still remain unknown despite our best efforts. In both classification problems, a single rotation of one building block in the unit cell can be sufficient to change class (Fig.~\ref{fig:modescaling}(c)). Hence, the boundary between classes C and I is sharp and sensitive to minimal perturbations as in the case of metamaterial M1 (Fig.~\ref{fig:smiley_designspace}(c)).

If the rules are unknown, the classification of this metamaterial requires the determination of $M(n)$---via rank-revealing QR factorization \cite{hong1992rank}---as function of the number of unit cells $n$, which is computationally demanding. For $k\times k$ unit cells, the time it takes to compute this brute-force classification scales nearly cubically with input size $k^2$. In contrast, classification with NNs \ch{scales linearly with input size and is readily parallelizable. In practice this makes NNs invariant to input size due to computational overhead} (see SM~\footnote{See Supplemental Material at [URL will be inserted by publisher] for a more detailed description of the computational time comparison.}). Hence a trained NN allows for much more time-efficient exploration of the design space.

To train our NNs, we generate labeled data through Monte Carlo sampling the design space to generate $5\times 5$ unit cells designs and explicitly calculate $M(n)$ for $n\in\{2, 3, 4\}$ to determine the classification. We do this for a range of $k\times k$ unit cells with $3\leq k \leq 8$. We focus on $5\times 5$ but the results hold for other unit cell sizes (see SM~\footnote{See Supplemental Material at [URL will be inserted by publisher] for CNN results of more unit cell sizes.}). The generated data is subsequently split into training (85\%) and test (15\%) sets. As our designs are spatially structured and local building block interactions drive compatible deformations, we ask whether convolutional
neural networks (CNNs) are able to distinguish between class C and I. The input of our CNNs are pixelated representations of our designs. This approach facilitates the identification of neighboring building blocks that are capable of compatible deformations (see SM~\footnote{See Supplemental Material at [URL will be inserted by publisher] for a more detailed description of the pixel representation.}). The CNNs are trained using 10-fold stratified cross-validation. Crucially, we use a balanced training set, where the proportion of class I has been randomly undersampled such that classes C and I are equally represented (see SM~\footnote{See Supplemental Material at [URL will be inserted by publisher] for more details about the training and test sets for each metamaterial.}).

Despite the complexity of the classification problems, we find that the CNNs perform very well (Tab.~\ref{tab:confusion matrices}). In particular, the CNNs 
correctly classify most class C unit cells as class C, and most class I unit cells as class I.
However, the test set is likely to contain few examples of class I close to the C-I boundary, especially as C becomes more rare (Fig.~\ref{fig:smiley_designspace}(c), see SM~\ref{fn: undersample C-I}). Hence, whether our CNNs capture the complex boundary of C cannot be deduced from the test set alone.
In other words, the CNNs find the needles in the haystack but it remains unclear whether the needles are approximated finely (Fig.~\ref{fig:smiley_designspace}(c)) or coarsely (Fig.~\ref{fig:smiley_designspace}(d))~\footnote{We have observed from qualitative analysis of the 29 falsely classified C unit cells of M2.i that all unit cells appearAs our designs are spatially structured and local building block interactions drive compatible deformations, we ask whether convolutional
neural networks (CNNs) are able to distinguish between class C and I. to be close to C in design space.}.

\begin{figure}[b]
\centering
\includegraphics{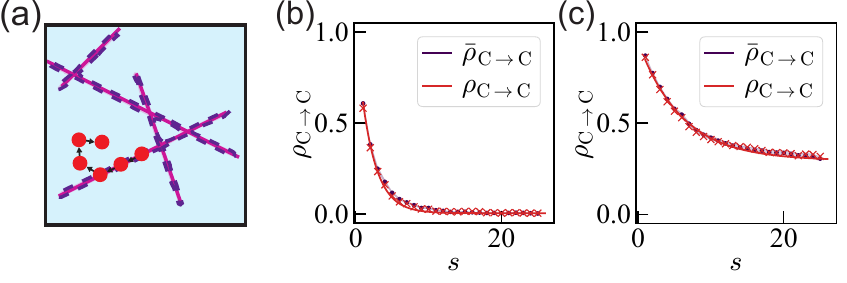}
\caption{\label{fig:random_walks}
(color online)
(a) Example of a 6-step random walk through design space (red dots) and sketch of the decision boundary of trained CNNs that has learned the combinatorial rules (purple dashed line).
(b) Probabilities to remain in true and predicted class C under random walks of $s$ steps, $\rho_{\text C \to \text C}(s)$ (red crosses) and fold-averaged $\langle \bar{\rho}_{\text C \to \text C}\rangle (s) $ (purple circles) with standard deviation (purple area), for classification (i) (left) and (ii) (right). 
The red continuous line is a least-squares fit to $\rho_{\text C \to \text C}(s)$ using Eq.~\eqref{eq: PtoP analytic}.
}
\end{figure}

\textit{Combinatorial structure}\textemdash
To probe the shape of both the true set of C configurations and the set of classified C configurations, we
start from a true class C configuration, perform random walks in configuration space, and at each step
probe the probabilities to be in the set of true class C (Fig.~\ref{fig:random_walks}(a)).
We randomly change the orientation of a single random building block
at each step $s \mapsto s + 1$ and average over 1000 realizations (see SM~\cite{fn:random_walks})
The probability to remain in true class C, $\rho_{\text C \to \text C}(s)$, decreases with $s$ and saturates to the class C volume fraction $\beta$ for classification (i) and (ii) (Fig.~\ref{fig:random_walks}(b)).
We note that we can fit this decay by a simple model, where
we assume
that subspace C is highly complex, so that the probabilities to
leave it are uncorrelated. For every step, there is a chance $\alpha$ to remain C. Once in class I, we assume any subsequent steps are akin to uniformly probing the design space such that the probability to become C is equal to the C volume fraction $\beta$.
Thus the probability to become C can be modeled as
\begin{equation}
    \rho_{\text C \to \text C}(s) = \alpha^s + \beta \left(1 - \alpha^{s-1}\right).
    \label{eq: PtoP analytic}
\end{equation}
The uncorrelated nature of the steps are consistent with a random needle structure (Fig.~\ref{fig:smiley_designspace}(c)), where the coefficient $\alpha \times 4^{5\times 5}$ corresponds to the average dimensionality of the needles and $\beta$ corresponds to their volume fraction. We can interpret $\alpha$ as the probability to not break the combinatorial rules when we randomly rotate a building block.

To see whether the CNNs are able to capture these key features of space C,
we repeat our random walk procedure using the CNNs' classification instead, starting from true and classified C configurations, and obtain the probability $\bar{\rho}_{\text C \to \text C}(s)$. The decay of the fold-averaged $\langle \bar{\rho}_{\text C \to \text C} \rangle(s)$ closely matches that of the true class C for classification problems (i) and (ii)
(Fig.~\ref{fig:random_walks}(b, c)). By fitting the predicted probability $\bar{\rho}_{\text C \to \text C}(s)$ for each fold to Eq.~\eqref{eq: PtoP analytic}, using measurements of the CNN's predicted volume fraction $\bar{\beta}$ over the test set to constrain the fit, we obtain the fold-averaged dimensionality $\bar{\alpha}$. For classification (i) we find $\bar{\alpha} \approx0.632 \pm 0.001$ closely matches the true $\alpha \approx 0.612 \pm 0.001$. In practice, $\alpha$ corresponds to the fraction of building blocks that are outside the relevant combinatorial strip.
Using a simple counting argument, we find good agreement with the lower-bound of $\alpha \simeq 3/5$ (see SM~\cite{fn:random_walks}). 
Similarly, for classification (ii) we find $\bar{\alpha} \approx 0.8514 \pm 0.0005$ closely matches $\alpha \approx 0.846 \pm 0.002$.
Our results thus demonstrate that CNNs successfully capture on average the  complex local shape of the combinatorial space C.
Even though during learning the algorithm 'sees' very few class I unit cells that are close the C-I boundary, the decision boundary still captures on average the sparsity and fine structure of the class C subset. Thus we conclude that the CNNs infer the combinatorial rules (Fig.~\ref{fig:smiley_designspace}(c)), rather than interpolate the shape in high dimensional design space (Fig.~\ref{fig:smiley_designspace}(d)).
In other words, CNNs are able not only to capture accurately the volume fraction of the needles, but also to finely distinguish between needle and hay.

\begin{figure}[b!]
\centering
\includegraphics{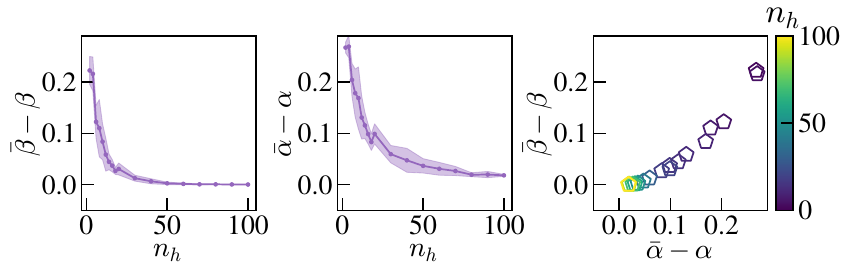}
\caption{\label{fig:beta_alpha_stripmodes}(color online) 
(a) Difference between predicted class C volume $\bar{\beta} (n_h)$ and true class C volume $\beta$ as a function of number of hidden neurons $n_h$ shows that $\bar{\beta}(n_h)$
    approaches $\beta$ for increasing $n_h$.
(b) Difference between predicted dimensionality $\bar{\alpha}(n_h)$ and true dimensionality $\alpha$ obtained through least-squares fits to Eq.~\eqref{eq: PtoP analytic} as a function of the number of hidden neurons $n_h$ shows that $\bar{\alpha}(n_h)$ approaches $\alpha$ for increasing $n_h$.
 (c) Scatter plots of class volumes $\bar{\beta}(n_h) - \beta$ versus dimensionality $\bar{\alpha}(n_h) - \alpha$ shows that the latter asymptotes later than the former ($n_h$ indicated by colorbar). We use CNNs with a single convolution layer of $20$ $2\times 2$ filters, which are spatially offset with respect to the unit cell and subsequently flattened. The flattened feature maps are fully-connected to a layer of $n_h$ hidden neurons, which itself is fully-connected to two output neurons that correspond to class C and I. The CNNs are systematically trained using 10-fold stratified cross-validation for varying number of hidden neurons $n_h$.
 }
\end{figure}

\textit{Volume before structure}\textemdash
But what happens with smaller CNNs?
We focus on classification (i) 
and probe how well our CNNs\ch{---which consist of a single 20 filters convolution layer, a single $n_h$ neurons hidden layer, and a two neurons output layer---}capture
the sparsity and structure of class C. First we compare their true and predicted volumes $\beta$ and $\bar{\beta}(n_h)$ as a function of the number of hidden neurons $n_h$.
The CNNs' predicted class C volume fraction $\bar{\beta}$ approaches the true class C volume fraction $\beta$ as the number of hidden neurons $n_h$ increases sufficiently, despite their balanced training set
(Fig.~\ref{fig:beta_alpha_stripmodes}(a))~\footnote{We note that $\mathrm{BA}(n_h)$ increases in conjunction with $\bar{\beta}(n_h)$, see Supplemental  Material at [URL will be inserted by publisher].}.
Next we compare the true and predicted dimensionality $\alpha$ and $\bar{\alpha}(n_h)$. While for small values of $n_h$, $\bar{\alpha}$ overestimates $\alpha$, $\bar{\alpha}$ closely matches $\alpha$ for large $n_h$ (Fig.~\ref{fig:beta_alpha_stripmodes}(b)).
For small number of hidden neurons $n_h$, the CNNs overestimate both the probability to remain in class C and the rarity of class C; in other words, small CNNs coarsen the complex shape of C (Fig.~\ref{fig:smiley_designspace}(c)). As seen above, for larger number of hidden neurons $n_h$ both the probability and rarity of C are closely approximated, thus large CNNs finely capture the complex shape of C (Fig.~\ref{fig:smiley_designspace}(d)).

Strikingly, we observe that the predicted class C volume $\bar{\beta}$ more quickly reaches its asymptotic value than the
dimensionality $\bar{\alpha}$. To see this, we plot 
$\bar{\beta}(n_h) - \beta$ versus  $\bar{\alpha}(n_h) - \alpha$, which demonstrates that after $\bar{\beta}$ closely approximates $\beta$, increasing the number of
hidden neurons $n_h$ improves $\bar{\alpha}(n_h)$ towards its asymptote $\alpha$ (Fig.~\ref{fig:beta_alpha_stripmodes}(c))---
this observation is also present for other unit cell sizes, see SM~\footnote{See Supplemental Material at [URL will be inserted by publisher] for measurements of $\bar{\alpha}$ and $\bar{\beta}$ of classification problem (i) for more unit cell sizes.}.
Thus, further increasing the size of the CNN beyond the point of marginal gain of test set performance results in a significantly more closely captured fine structure of C.
In other words, to correctly capture the average dimensionality of the needles requires more neurons than to capture their volume.

\textit{Discussion}\textemdash
NNs are  known to be universal approximators~\cite{hornik1989multilayer} and efficient classifiers. They often generalize well when the training data samples representative portions of the input space sufficiently, even for non-smooth~\cite{pmlr-v89-imaizumi19a} or noisy data~\cite{rolnick2017deep}. As combinatorial problems are sharply delineated and severely class-imbalanced, one expects that the fine details of an undersampled complex boundary would be blurred by NNs. Surprisingly, we have shown that CNNs will closely approximate such a complex combinatorial structure, despite being trained on a sparse training set. We attribute this to the underlying set of rules which govern the complex space of compatible configurations---in simple terms, the CNN learns the combinatorial rules, rather than the
geometry of design space, which is the complex result of those rules~\footnote{We expect NNs to work beyond combinatorial metamaterials for a wide range of combinatorial problems in physics, such as spin-ice. These combinatorial rules in such problems can typically be translated to matrix operations, NNs naturally capture such matrix operations, and therefore we expect them to perform well.}.

Recognizing NNs' ability to learn these rules from a sparse representation of the design space opens new strategies for design. 
For instance, our CNNs could be readily used as surrogate models within a design algorithm to save computational time.
Alternatively, 
one could instead devise a design algorithm based on generative adversarial NNs~\cite{goodfellow2014generative} or variational auto-encoders~\cite{kingma2013auto}. It is an open question whether and how such 
generative models
could successfully leverage the 
learning of combinatorial rules~\cite{bengio2021machine}.

Our work shows that metamaterials provide a compelling avenue for machine learning combinatorial problems, as they are straightforward to simulate yet exhibit complex combinatorial structure (Fig.~\ref{fig:smiley_designspace}(c)). More broadly, applying neural networks to combinatorial problems opens many exciting questions. What is the relation between the complexity of the combinatorial rules and that of the networks? Can unsolved combinatorial problems be solved by neural networks? Can neural networks learn size-independent combinatorial rules? Conversely, can these problems help us understand why neural networks work so well~\cite{zhang2021understanding}? Can they provide insight in how to effectively overcome strong data-imbalance~\cite{johnson2019survey}? We believe combinatorial metamaterials are well suited to answer such questions.

\begin{acknowledgments}
\textit{Data availability statement.}\textemdash The code supporting the findings reported in this paper is publicly available on GitHub~\footnote{See \url{https://uva-hva.gitlab.host/published-projects/CombiMetaMaterial} for code to calculate zero modes}\footnote{See \url{https://uva-hva.gitlab.host/published-projects/CNN_MetaCombi} for code to train and evaluate convolutional neural networks.}---the data on Zenodo~\cite{Zenodo_MetaCombi,  Zenodo_CNN}.

\textit{Acknowledgements.}\textemdash We thank David Dykstra, Marc Serra-Garcia, Jan-Willem van de Meent, and Tristan Bereau for discussions. This work was carried out on the Dutch national e-infrastructure with the support of SURF Cooperative. C.C. acknowledges funding from the European Research Council under Grant Agreement 852587. 
\end{acknowledgments}

\bibliography{bibliography}

\newpage
\appendix

\setcounter{figure}{0}
\renewcommand{\thefigure}{A\arabic{figure}}%

\setcounter{table}{0}
\renewcommand{\thetable}{A\arabic{table}}%

\setcounter{equation}{0}
\renewcommand{\theequation}{A\arabic{equation}}%

\subsection{Floppy and frustrated structures}
In this section, we discuss in more detail the metamaterial M1 of Fig.~1. We first derive the design rules that lead to floppy structures, then we discuss the rarity of such structures.

\subsubsection{Design rules for floppy structures}
Here we provide a brief overview of the rules that lead to floppy structures for the combinatorial metamaterial M1 of Fig.~1. The three-dimensional building block of this metamaterial can deform in one way that does not stretch any of the bonds: it has one zero mode (see \cite{coulais2016combinatorial} for details of the unit cell). In two dimensions, there are two orientations of the building block that deform differently in-plane. We label these two orientations as green/red and white (Fig.~1(a)).

We can formulate a set of rules for configurations of these building blocks in two dimensions. Configurations of only green/red building blocks or white building blocks deform compatibly (C): the configuration is floppy. A single horizontal or vertical line of white building blocks in a configuration filled with green/red building blocks also deforms compatibly. More lines (horizontal or vertical) of white blocks in a configuration filled with green/red blocks deform compatibly if the building block at the intersection of the lines is of type green/red (Fig.~1(b)). 

In summary, we can formulate a set of rules:
\begin{enumerate}[i]
    \item All white building blocks need to be part of a horizontal or vertical line of white building blocks.
    \item At the intersection of horizontal and vertical lines of white building blocks there needs to be a green/red building block.
\end{enumerate}
If these rules are met in a configuration, the configuration will be floppy (C). A single change of building block is sufficient to break the rules, creating an incompatible (I) frustrated configuration (Fig.~1(b)). 

\subsubsection{Rarity of floppy structures}
\begin{figure}[b]
    \centering
    \includegraphics{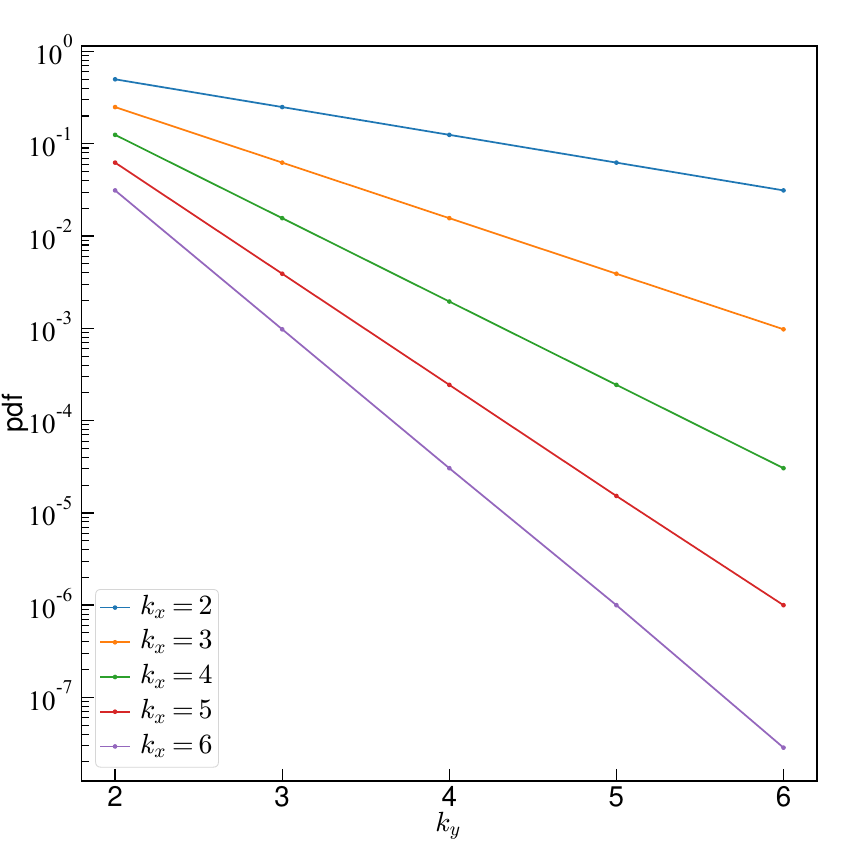}
    \caption{Probability density function (pdf) of $k_x \times k_y$ class C configurations.}
    \label{fig:smiley_Crarity}
\end{figure}

Here we show how the rarity of class C depends on the size of the $k_x \times k_y$ configuration. To show this, we simulate configurations with varying $k_x, k_y \in \{2, 3, 4, 5, 6\}$. The size of the design space grows exponentially as $2^{k_x k_y}$, yet the fraction of class C configurations decreases exponentially with unit cell size (Fig.~\ref{fig:smiley_Crarity}). Thus the number of C configuration scale with unit cell size at a much slower rate than the number of total configurations. For large configuration size, the number of C configurations is too small to create a sufficiently large class-balanced training set to train neural networks on. 

\subsection{Zero modes in combinatorial metamaterials}
In this section, we present theoretical and numerical results at the root of the classification of zero modes in the combinatorial metamaterial M2 of Fig.~2. We first derive the zero modes of the building block, then we postulate a set of rules for classification (i) of unit cells. Finally, we provide numerical proof of those rules.
\subsubsection{\label{SM: Building Block}Zero Modes of the Building Block}
\begin{figure}[b]
    \centering
    \includegraphics[width=0.25\textwidth]{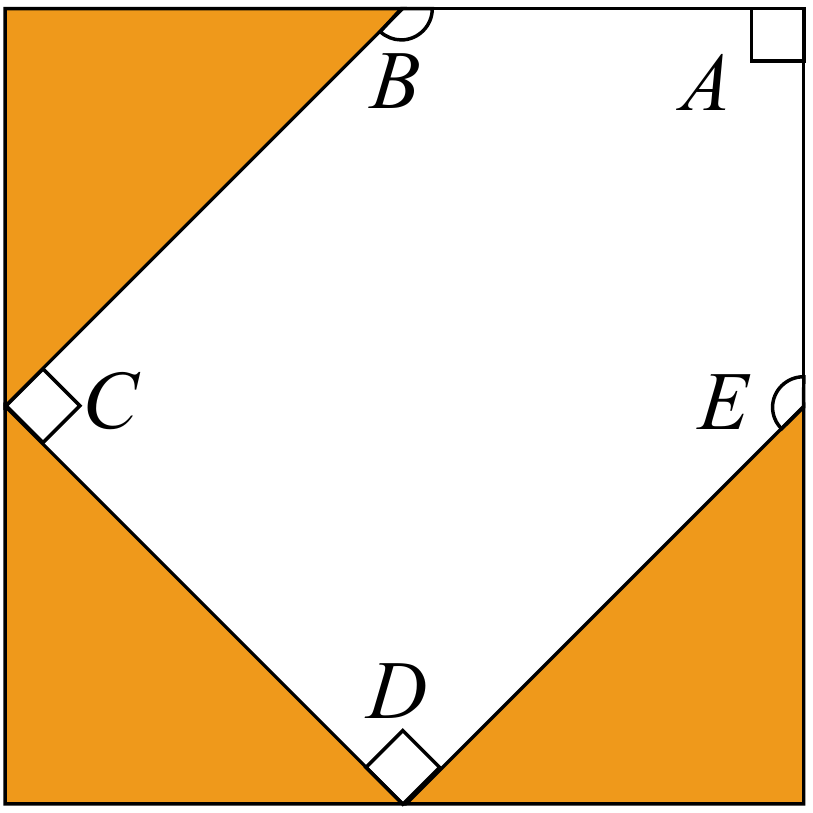}
    \caption{Schematic real space representation of the building block. $A$, $B$, $C$, $D$, and $E$ label the five corners that can change angle under zero-energy deformations.}
    \label{fig:building block ABCDE}
\end{figure}

The fundamental building block is shown schematically in Fig.~\ref{fig:building block ABCDE}. Each black line represents a rigid bar, while vertices can be thought of as hinges; the 11 bars are free to rotate about the 8 hinges in 2 dimensions. The colored triangles form rigid structures, \textit{i.e.} they will not deform. From the Maxwell counting~\cite{maxwell1864calculation} we obtain
$    N_{zm} = 2\cdot 8 - 11 - 3 = 2,$
where the 3 trivial zero modes in 2 dimensions, translation and rotation, are subtracted such that $N_{zm}$ is the number of zero modes of the building block. The precise deformation of these two zero modes can be derived from the geometric constraints of the building block.

To derive the zero modes to linear order, we note that they preserve the length of all bars, such that the modes can be characterized by the hinging angles of the bar. Let $A, B, C, D$, and $E$ denote these angles. Going around the loop $ABCDE$, the angles add up to $3 \pi$:
\begin{equation}
    A + B + C + D + E = 3 \pi.
    \label{eq: trig 1}
\end{equation}
Next, we expand the angles from their rest position to linear order:
\begin{equation}
    \begin{tabular}{ccc}
        $A=\frac{\pi}{2} + \alpha$, & $B = \frac{3 \pi}{4} + \beta$,  & $C = \frac{\pi}{2} + \gamma$, \\
       $ D = \frac{\pi}{2} + \delta$, & $E=\frac{3 \pi}{4} + \epsilon$. \\ 
    \end{tabular}
\end{equation}
Then, from the condition that the bars cannot change length, we obtain
\begin{equation}
    1 - \cos{(A)} = 3-2 \cos{(C)} - 2\cos{(D)} +2 \cos{(C+D)}, \,
    \label{eq: trig 2}
\end{equation}
and 
\begin{equation}
    \sin{(D)} - \frac{\sin{(D+E)}}{\sqrt{2}} = \sin{(C)} - \frac{\sin{(C+B)}}{\sqrt{2}}.
    \label{eq: trig 3}
\end{equation}
Up to first order in $\alpha, \beta, \gamma, \delta, \epsilon$, equations \eqref{eq: trig 2} and \eqref{eq: trig 3} can be rewritten as:
\begin{equation}
    \alpha = 2 \gamma + 2 \delta,
    \label{eq: trig 2 linear}
\end{equation}
\begin{equation}
    \delta + \epsilon = \beta + \gamma.
    \label{eq: trig 3 linear}
\end{equation}

Together with the loop condition \eqref{eq: trig 1}, we obtain a set of three equations which express $\alpha, \delta$ and $\epsilon$ in $\beta$ and $\gamma$:
\begin{equation}
    \begin{pmatrix}
        \alpha \\
        \delta \\
        \epsilon
    \end{pmatrix} = \begin{pmatrix}
        -2  &   -2  \\
        -1  &   -2  \\
        2   &   3   \\
    \end{pmatrix} \begin{pmatrix}
        \beta \\
        \gamma \\
    \end{pmatrix}.
\end{equation}
This demonstrates that we can choose the two parameters $\beta$ and $\gamma$ arbitrarily, while still satisfying equations \eqref{eq: trig 1}, \eqref{eq: trig 2 linear} and \eqref{eq: trig 3 linear}, consistent with the presence of two zero modes.

We now choose the basis of the zero modes such that the first zero mode is the deformation of the square BCDE, such that $\alpha = 0$. This leads to the well-known counter-rotating squares (CRS) mode~\cite{grima2005auxetic, coulais2018characteristic} when tiling building blocks together. Thus we choose the basis
\begin{equation}
    \begin{pmatrix}
        \beta \\
        \gamma
    \end{pmatrix} = M_{CRS} \begin{pmatrix}
        -1 \\
        1
    \end{pmatrix} + M_{D} \begin{pmatrix}
        3 \\
        -1
    \end{pmatrix}.
\end{equation}
$M_{CRS}$ is the amplitude for the counter-rotating squares mode, while $M_{D}$ is the amplitude of the mode that does change corner $A$. We refer to this mode as the diagonal mode.

By tiling together the building block in different orientations, we can create $4^{k^2}$ size $k \times k$ unit cells. 
These unit cells --- and metamaterials built from them --- may have more or less zero modes than the constituent building blocks, depending on the number of states of self-stress.
Previous work on $2 \times 2$ unit cells showed that each unit cell could be classified based on the number of zero modes~\cite{bossart2021oligomodal}. Here, we consider the previously unexplored cases of $3\times 3$ up to $8\times 8$ square unit cells.

\subsubsection{Rule-based classification of unit cells}
Unit cells are classified based on the number of zero modes $M(n)$ for $n\geq 2$ as either class I or class C as described in the main text. Here we formulate a set of empirical rules that distinguishes class I unit cells from class C unit cells for classification (i).

\begin{figure}[b]
    \centering
    \includegraphics{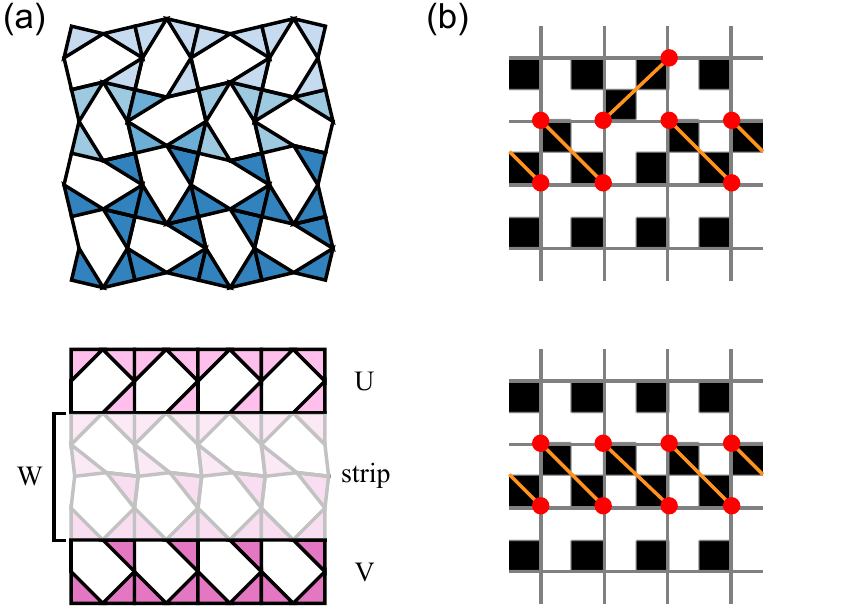}
    \caption{Schematic and pixel representation of modes in a $4\times 4$ unit cell. (a) Schematic deformation of counter-rotating squares mode (top unit cell, blue) and a strip mode (bottom unit cell, pink). The strip mode spans the entire area of the strip (white) of width $W=2$, while the areas U and V do not deform. (b) Respective pixel representations of the left unit cells. Paired unit cells are highlighted through red dots connected by orange lines. Note that the top unit cell does not contain a strip that meets the strip mode rules, while the bottom unit cell does.}
    \label{fig:CRS_LM_Schem_PixRep}
\end{figure}

Any finite configuration of building blocks, no matter the orientation of each block, supports the counter-rotating squares (CRS) mode with open boundary conditions, where all building blocks will deform with $M_{CRS}\neq 0$ and $M_{D}=0$. They must all have equal magnitude $|M_{CRS}|$, but alternate in sign from building block to building block in a checkerboard pattern, similar to the ground state of the anti-ferromagnetic Ising model on a square lattice. An arbitrary configuration in the real space representation, and the CRS mode of that configuration in the directed graph representation, are shown in Fig.~\ref{fig:CRS_LM_Schem_PixRep}(a).

However, precisely because the building block supports another mode, there could in principle be other collective modes than the CRS mode in any given configuration. We have observed that class C unit cells have a specific structure, which we refer to as a strip mode. A strip mode spans the unit cell periodically in one direction, such that the total number of zero modes for a configuration of $n\times n$ tiled unit cells grows linearly with $n$. 

The pattern of deformations for these modes consists of two rectangular patches of building blocks with CRS modes (where $M_{D}=0$ for every building block) --- potentially of different amplitude --- separated by a strip of building blocks (the \textit{strip}) that connects these patches, where $M_{D} \neq 0$. A unit cell configuration with a strip mode, which consists of building blocks in a strip of block-width $W=2$ that deform with $M_{D}\neq 0$, and building blocks in the two areas outside of the strip, U \& V, that do not deform, is shown in Fig.~\ref{fig:CRS_LM_Schem_PixRep}(a). Note that the CRS mode can always be freely added or subtracted from the total configuration. 

\renewcommand{\labelenumi}{\roman{enumi}}
\newcounter{rules}
\begin{enumerate}
    \item We conjecture that the presence of a strip mode is a necessary and sufficient condition for a unit cell to be of class C.
    \setcounter{rules}{\value{enumi}}
\end{enumerate}

We verify (i) below. Moreover, we now conjecture a set of necessary and sufficient conditions on the configuration of the strip that lead to a strip mode. Underlying this set of conditions is the notion of \textit{paired} building blocks: neighboring blocks that connect with their respective $A$ corners, or equivalently, blocks that have their black pixels in the same plaquette in the pixel representation, see Fig.~\ref{fig:CRS_LM_Schem_PixRep}(b). Depending on the orientation of the paired building blocks, pairs of these blocks are referred to as horizontal, vertical or diagonal pairs. The set of conditions to be met within the strip to have a horizontal (vertical) strip mode can be stated as follows:

\begin{enumerate}
    \setcounter{enumi}{\value{rules}}
    \item Each building block in the strip is paired with a single other neighboring building block in the strip.
    \item Apart from horizontal (vertical) pairs, there can be either vertical (horizontal) or diagonal pairs within two adjacent rows (columns) in the strip, never both.
\end{enumerate}
Consider the unit cells of Fig.~\ref{fig:CRS_LM_Schem_PixRep}, the top unit cell has multiple paired building blocks, but contains no horizontal (or vertical) strip where every block is paired. Conversely, the bottom unit cell does contain a strip of width $W=2$ blocks where every block is paired to another block in the strip. Consequently, the bottom unit cell obeys the rules and supports a strip mode, while the top unit cell does not.

Each indivisible strip of building blocks for which these conditions hold, supports a strip mode. For example, if a unit cell contains a strip of width $W=2$ which obeys the rules, but this strip can be divided into two strips of width $W=1$ that each obey the rules, then the width $W=2$ strip supports two strip modes, not one. 

We refer to (i) as the strip mode conjecture, and (ii) and (iii) as the strip mode rules. 
We now present numerical evidence that supports these rules.

\subsubsection{Numerical evidence for strip mode rules}
The conjecture and rules (i)-(iii) stated in the previous section can be substantiated through numerical simulation. To do so, we determine the class of randomly picked unit cells.

To assess the rules, a large number of square unit cells are randomly generated over a range of sizes $k\in \{3,4,5,6,7,8\}$. For each unit cell configuration, $n_x\times n_y$ metamaterials, composed by tiling of the unit cells, are generated over a range of $n_x=n_y=n\in \{1, 2, 3, 4\}$ for $k\leq 4$. From $k\geq 6$ onward, the $1\times 1$ configuration is generated, as well as $n_x \times 2$ and $2 \times n_y$ configurations with $n_x, n_y \in \{2, 3, 4\}$ to save computation time.

The rigidity, or compatibility, matrix $R$ is constructed for each of these configurations, subsequently rank-revealing QR factorization is used to determine the dimension of the kernel of $R$. This dimension is equivalent to the number of zero modes of the configuration, $M(n)$ is then equal to this number minus the number of trivial zero modes: two translations and one rotation.

From the behavior of $M(n)$ as a function of $n$, we define the two classes: I and C. In Class I $M(n)$ saturates to a constant for $n \geq 2$, thus class I unit cells do not contain any strip modes. Note that they could still contain additional zero modes besides the CRS mode. In Class C $M(n)$ grows linearly with $n$ for $n \geq 2$, therefore class C unit cells could support a strip mode \footnote{There is a small and exponentially decreasing portion of unit cells that requires to calculate $M(n)$ with $n=5$ and $6$ to determine whether they belong to class I or C. We leave these out of consideration in the training data to save computational time.}. Moreover, if conjecture (i) is true, the number of strip modes supported in the class C configuration should be equivalent to the slope of $M(n)$ from $n\geq 2$ onward.

In class I, $M(n)$ is constant for sufficiently large $n$, thus class I unit cells do not contain any strip modes. Note that they could still contain additional zero modes besides the CRS mode. In class C $M(n)$ grows linearly with $n$ for sufficiently large $n$, therefore class C unit cells could support a strip mode. Moreover, if conjecture (i) is true, the number of strip modes supported in the class C configuration should be equivalent to the slope of $M(n)$ for sufficiently large $n$.

\begin{figure}[b]
    \centering
    \includegraphics{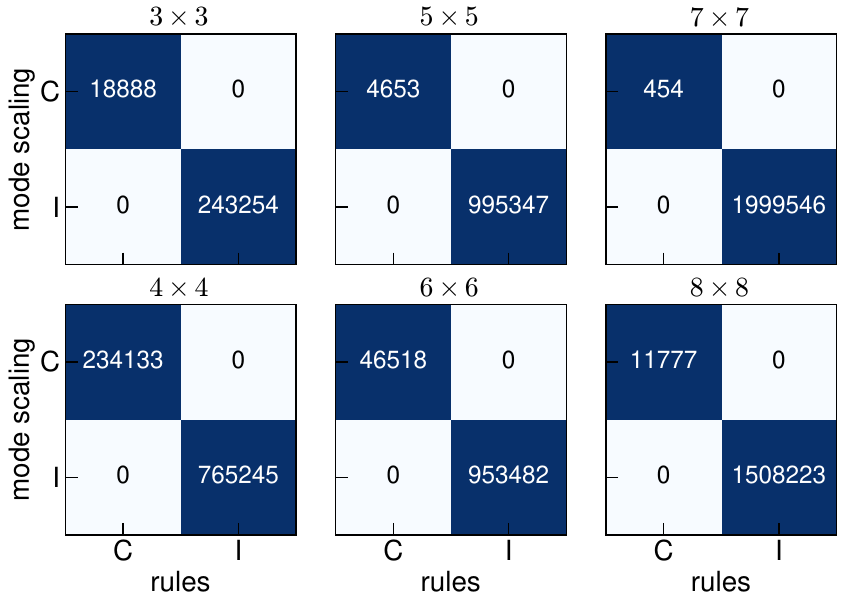}
    \caption{Confusion matrices for classification based on mode scaling in comparison to classification based on rules (i)-(ii). The $k \times k$ unit cell size is indicated on top of each matrix.}
    \label{fig:pdf_rules_ABC}
\end{figure}

To test conjecture (i) and the strip mode rules (ii) and (iii), we check for each generated unit cell if it contains a strip that obeys the strip mode rules. This check can be performed using simple matrix operations and checks~\footnote{See \url{https://github.com/Metadude1996/CombiMetaMaterial} for code to check the rules.}. If (ii)-(iii) are correct, the number of indivisible strips that obey the rules within the unit cell should be equal to the slope of $M(n)$ for class C unit cells, and there should be no strips that obey the rules in class I unit cells. Simulations of all possible $k=3$ unit cells, one million $k=4, 5, 6$ unit cells, two million $k=7$ unit cells, and 1.52 million $k=8$ unit cells show perfect agreement with the strip mode rules for unit cells belonging to either class I or C, see Fig.~\ref{fig:pdf_rules_ABC}.  Consequently, numerical simulations provide strong evidence that the strip mode rules as stated are correct.

\subsection{Constructing and Training Convolutional Neural Networks for metamaterials}

In this section, we describe in detail how we construct and train our convolutional neural networks (CNNs) for classifying unit cells into class I and C. 
We first transform our unit cells to a CNN input, secondly we establish the architecture of our CNNs. Next, we obtain the training set, and finally we train our CNNs.

\subsubsection{Pixel Representation}
To feed our design to a neural network, we need to choose a representation a neural network can understand. Since we aim to use convolutional neural networks, this representation needs to be a two-dimensional image. For our classification problem, the presence or absence of a zero mode ultimately depends on compatible deformations between neighboring building blocks. As such, the representation we choose should allow for an easy identification of the interaction between neighbors.

In addition to being translation invariant, the classification is rotation invariant. While we do not hard code this symmetry in the convolutional neural network, we do choose a representation where rotating the unit cell should still yield a correct classification. For example, this excludes a representation where each building block is simply labeled by a number corresponding to its orientation. For such a representation, rotating the design without changing the numbers results in a different interplay between the numbers than for the original design. Thus we cannot expect a network to correctly classify the rotated design.

\begin{figure}[b]
    \centering
    \includegraphics{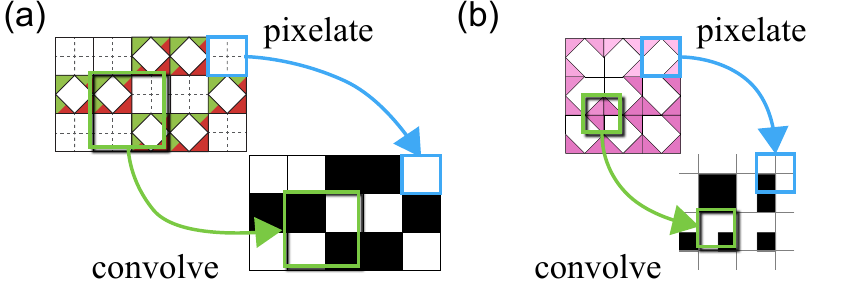}
    \caption{Unit cell designs of the combinatorial metamaterials in Fig.~1 (a) and Fig.~2 (b) and their respective pixel representations. The blue squares indicates how the building blocks are transformed to pixels, the green squares show which part of the unit cell is convolved by the first convolution layer.}
    \label{fig:pixelreps}
\end{figure}

For both metamaterials, we introduce a \textit{pixel} representation. We represent the two building blocks of metamaterial featured in Fig.~1 as either a black pixel (1) or a white pixel (0) (Fig.~\ref{fig:pixelreps}(a)). A $k_x \times k_y$ unit cell thus turns into a $k_x \times k_y$ black-and-white image.

Likewise, we introduce a \textit{pixel} representation for the metamaterial M2 of Fig.~2 which naturally captures the spatial orientation of the building blocks, and emphasizes the interaction with neighboring building blocks. In this representation, each building block is represented as a $2\times 2$ matrix, with one black pixel (1) and three white (0) pixels, see Fig.~\ref{fig:pixelreps}(b). The black pixel is located in the quadrant where in the bars-and-hinges representation the missing diagonal bar is. Equivalently, this is the quadrant where in the directed graph representation the diagonal edge is located. Moreover, in terms of mechanics, this quadrant can be considered floppy, while the three others are rigid.

This representation naturally divides the building blocks into $2\times 2$ \textit{plaquettes} in which paired building blocks are easily identified, see Fig.~\ref{fig:pixelreps}(b). Building blocks sharing their black pixel in the same plaquette are necessarily paired, and thus allow for deformations beyond the counter-rotating squares mode. Note that this includes diagonally paired building blocks as well. By setting the stride of the first convolution layer to $(2, 2)$, the filters only convolve over the plaquettes and not the building blocks, which do not contain any extra information for classification.

\subsubsection{CNN architecture details}
To classify the unit cells into class I and C, we use a convolutional neural network (CNN) architecture. We first discuss the architectures used to obtain the results of Tab.~1. Then we discuss the architecture used to obtain the results of Fig.~4.

For the metamaterial M1 of Fig.~1, the CNN consists of a single convolution layer with 20 $2\times 2$ filters with bias and ReLu activation function. The filters move across the input image with stride $(1, 1)$ such that all building block interactions are considered. Subsequently the feature maps are flattened and fully-connected to a hidden layer of 100 neurons with bias and ReLu activation function. This layer subsequently connected to 2 output neurons corresponding to C and I with bias and softmax activiation function. The input image is not padded. Since a network of this size was already able to achieve perfect performance, we saw no reason to go to a bigger network.

For the metamaterial M2 of Fig.~2 and classification problem (i) 
we first periodically pad the input image with a pixel-wide layer, such that a $2k \times 2k$ image becomes a $2k+2 \times 2k+2$ image.  This image is then fed to a convolutional layer, consisting of 20 $2\times 2$ filters with bias and ReLu activation function. The filters move across the input image with stride $(2, 2)$, such that the filters always look at the parts of the image showing the interactions between four building blocks (Fig.~\ref{fig:pixelreps}(b)). Subsequently the 20 $k+1 \times k+1$ feature maps are flattened and fully-connected to a hidden layer of 100 neurons with bias and ReLu activation function. This layer is then fully-connected to 2 output neurons corresponding to the two classes with bias and softmax activation function. From the hyperparameter grid search (see section \emph{CNN hyperparameter grid search details}) we noted that this $n_f$ and $n_h$ were sufficiently large for good performance.

For classification (ii) we again pad the input image with a pixel-wide layer. The CNN now consists of three sequential convolutional layers of increasing sizes 20, 80, and 160 filters with bias and ReLu activation function. The first convolution layer moves with stride $(2, 2)$. The feature maps after the last convolutional layer are flattened and fully-connected to a hidden layer with 1000 neurons with bias and ReLu activation function. This layer is fully-connected to two output neurons with bias and softmax activation function. This network is larger than for classification (i); we saw noticeable improvements over the validation set when we considered larger networks. This is most likely a result of the (unknown) rules behind classification (ii) being more complex.

The networks are trained using a cross-entropy loss function. This loss function is minimized using the Adam optimization algorithm~\cite{kingma2014adam}. This algorithm introduces additional parameters to set before training compared to stochastic gradient descent. We keep all algorithm-specific parameters as standard ($\beta_1=0.9$, $\beta_2=0.999$, $\epsilon=1\mathrm{e}-07$), and only vary the learning rate $\eta$ from run to run. The network for the classification problem of Fig.~1 uses a weighted cross-entropy loss function, where examples of C are weighted by a factor 200 more than examples of I.

To obtain the results of Fig.~4, we use the architecture of classification (i) and vary the number of neurons $n_h$ in the hidden layer. We keep the number of filters the same. To obtain this architecture, we performed a hyperparameter grid search, where we varied the number of filters $n_f$ of the convolution layer and the learning rate $\eta$ as well. The details are discussed in the section \textit{CNN hyperparameter grid search details}. The total number of parameters for this network with $n_f$ filters and $n_h$ neurons is \begin{equation}
    (4+1)n_f + ((k+1)^2 n_f+1) n_h + (n_h+1) 2.
    \label{eq: params CNN}
\end{equation}

\subsubsection{Training set details}
Each classification problem has its own training set. For the classification problem of Fig.~1, the networks are trained on a training set $D_t$ of size $|D_t|=27853$ that is artificially balanced 200-to-1 I-to-C. Classification problem (i) has a class balanced training set size of $|D_t|=793200$. Problem (ii) has a training set size of $|D_t|=501850$.
For the classification problems (i) and (ii), the class is determined through the total number of modes $M(n)$ as described in the subsection \textit{Numerical evidence for strip mode rules}. For the metamaterial M1 of Fig.~1, we determine the class through the rules as described in the section \emph{Floppy and frustrated structures}.

Since there is a strong class-imbalance in the design space, for the network to learn to distinguish between class I and C, the training set is class-balanced. If the training set is not class-balanced, the networks tend to learn to always predict the majority class. The training set is class-balanced using random undersampling of the class I designs. For problem (i), with the strongest class-imbalance, the number of class C designs is artificially increased using translation and rotation of class C designs. We then use stratified cross-validation over 10 folds, thus for each fold 90\% is used for training and 10\% for validation. The division of the set changes from fold to fold. To pick the best performing networks, we use performance measures measured over the validation set.

To show that our findings are robust to changes in unit cell size, we also train CNNs on classification problem (i) for different $k\times k$ unit cell sizes. The size of the training set $D_{\mathrm{t}}$ for each unit cell size $k$ is shown in Tab.~\ref{tab:CNNgridsearchdetails}. Increasing the unit cell size increases the rarity of C and the size of the design space. This leads a more strongly undersampled C-I boundary as we will show in the next section.

\begin{table}[b!]
\centering
\caption{\label{tab:CNNgridsearchdetails}Details of the hyperparameter grid search.}
\begin{ruledtabular}
\begin{tabular}{p{0.1\linewidth} p{0.4\linewidth} p{0.4\linewidth}}
 $k$ & size of $D_{t}$ & size of $D_{\mathrm{test}}$\\ \colrule
 3 & 31180 & 39321 \\
 4 & 397914 & 150000  \\
 5 & 793200 & 149980 \\
 6 & 1620584 & 150000  \\
 7 & 292432 & 600000 \\
 8 & 1619240 & 144000 \\
\end{tabular}
\end{ruledtabular}
\end{table} 

\subsubsection{Sparsity of the training set}
\begin{figure}[b]
    \centering
    \includegraphics{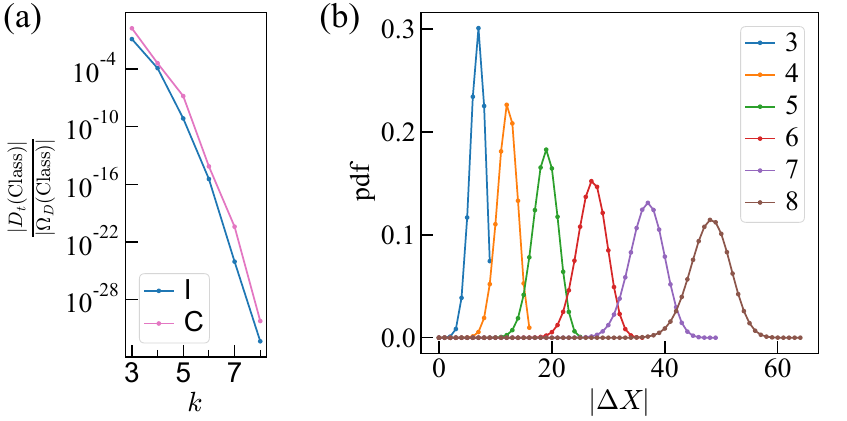}
    \caption{Training set details for classification problem (i) of metamaterial M2. (a) Fraction of the total unit cells of class C that are in the training set. (b) Average absolute distance $|\Delta X|$ in number of building blocks between class C and class I unit cells in the training set.}
    \label{fig:trainset_details}
\end{figure}

To illustrate how sparse the training set is for classification problem (i), we divide the number of training unit cells per class, $|D_t(\mathrm{Class})|$ over the estimated total number of $k\times k$ unit cells of that class, $|\Omega_D(\mathrm{Class})|$. We estimate this number for class C through multiplying the volume fraction of class C $\beta$ in a uniformly generated set of unit cells with the total number of possible unit cells $|\Omega_D|=4^{k^2}$: $|\Omega_D(\mathrm{C})| \approx \beta |\Omega_D|$. Likewise, we determine the ratio for class I. The resulting ratio for class C and I is shown in Fig.~\ref{fig:trainset_details}(a). Clearly, for increasing unit cell size $k$, the class sparsity in the training set increases exponentially. Consequently, the neural networks get relatively fewer unit cells to learn the design rules bisecting the design space for increasing unit cell size. 

Moreover, the training set unit cells of different class are, on average, farther removed from one another for increasing unit cell size $k$. The distance between two unit cells $|\Delta X|$ is defined as the number of building blocks with a different orientation compared to their corresponding building block at the same spatial location in the other unit cell. So two $k\times k$ unit cells can at most be $k^2$ building blocks removed from one another, if every single building block has a different orientation compared to its corresponding building block at the same spatial location in the other unit cell. Note that we only consider \textit{different} orientations in this definition, we do not define an additional notion of distance between orientations of building blocks.

By measuring the distance in number of different building block orientations $|\Delta X|$ between every class C to every class I unit cell, we obtain the probability density function of distance in number of different building blocks between two unit cells of different class in the training set, see Fig.~\ref{fig:trainset_details}(b). Consequently, if $k$ increases, the networks are shown fewer examples of unit cells similar to each other, but of different class. Thus the boundary between C and I is undersampled in the training set, with few I designs close to the boundary.

\subsubsection{CNN hyperparameter grid search details}
To see how convolutional neural network (CNN) size impacts classification performance, a hyperparameter grid search is performed. We focus on classification problem (i), which features a shallow CNN with a single convolution layer and single hidden layer as described in section \emph{CNN architecture details}. This search varied three hyperparameters: the number of filters $n_f$, the number of hidden neurons $n_h$, and the learning rate $\eta$. 
The number of filters $n_f$ runs from 2 to 20 in steps of 2, the number of hidden neurons $n_h$ first runs from 2 to 20 in steps of 2, then from 20 to 100 in steps of 10. The learning rate ranges from $\eta \in {0.0001, 0.001, 0.002, 0.003, 0.004, 0.005}$.
For each possible hyperparameter combination, a 10-fold stratified cross validation is performed on a class-balanced training set. Early stopping using the validation loss is used to prevent overfitting.

To create the results of Fig.~4, $n_f$ has been fixed to 20 since most of the performance increase seems to come from the number of hidden neurons $n_h$ after reaching a certain treshhold for $n_f$ as we will show in section \emph{Assessing the performances of CNNs}. The best $\eta$ is picked by selecting the networks with the highest fold-averaged accuracy over the validation set. 

\subsection{Assessing the performances of CNNs}
In this section, we describe in detail how we assess the performance of our trained convolutional neural networks (CNNs). 
We first quantify performance over the test set, then we define our sensitivity measure. Finally, we apply this sensitivity measure to the CNNs.

\subsubsection{Test set results}
After training the CNNs on the training sets, we test their performance over the test set. The test set consists of unit cells the networks have not seen during training, and it is not class-balanced. Instead, it is highly class-imbalanced, since the set is obtained from uniformly sampling the design space. In this way, the performance of the network to new, uniformly generated designs is fairly assessed. 

For the classification problem of metamaterial M1, the test set $D_{\mathrm{test}}$ has size $| D_{\mathrm{test}}| = 4915$. Classification problem (i) for metamaterial M2 has test set size $|D_{\mathrm{test}}| = 149982$. Problem (ii) for M2 has test set size $| D_{\mathrm{test}}| = 149980$. 

Precisely because the test set is imbalanced, standard performance measures, such as the accuracy, may not be good indicators of the actual performance of the network. There is a wide plethora of measures to choose from~\cite{hossin2015review}. To give a fair assessment of the performance, we show the confusion matrices over the test sets for the trained networks with the lowest loss over the validation set in Tab.~1.

\subsubsection{Varying the unit cell size}
To see how the size of the unit cell impacts network performance, we performed a hyperparameter grid search as described in section \emph{CNN hyperparameter grid search details} for $k \times k$ unit cells ranging from $3 \leq k \leq 8$. We focus on classification problem (i). The size of the test set $D_{\mathrm{test}}$ is shown in Tab.~\ref{tab:CNNgridsearchdetails}.

To quantify the performance of our networks in a single measure, we use the Balanced Accuracy:
\begin{align}
    \mathrm{BA} &=\left\langle \frac{1}{2} \left( \frac{V_\mathrm{TC}}{V_\mathrm{TC} + V_\mathrm{FI}} + \frac{V_\mathrm{TI}}{V_\mathrm{TI} + V_\mathrm{FC}} \right)\right\rangle \\
        &= \left\langle \frac{1}{2} \left( \mathrm{TCR} + \mathrm{TIR} \right) \right\rangle,
\end{align}
where $V_\mathrm{TC}$, $V_\mathrm{TI}$, $V_\mathrm{FC}$, and $V_\mathrm{FI}$ are the volumes of the subspaces true class C $\mathrm{TC}$, true class I $\mathrm{TI}$, false class C $\mathrm{FC}$, and false class I $\mathrm{FI}$ (Fig.~1(c, d)). We do not consider other commonly used performance measures for class-imbalanced classification, such as the $F_1$ score, since they are sensitive to the class-balance.

The $\mathrm{BA}$ can be understood as the arithmetic mean between the true class C rate $\mathrm{TCR}$ (sensitivity), and true class I rate $\mathrm{TIR}$ (specificity). As such, it considers the performance over all class C designs and all class I designs separately, giving them equal weight in the final score. Class-imbalance therefore has no impact on this score.

\begin{figure}[b]
    \centering
    \includegraphics{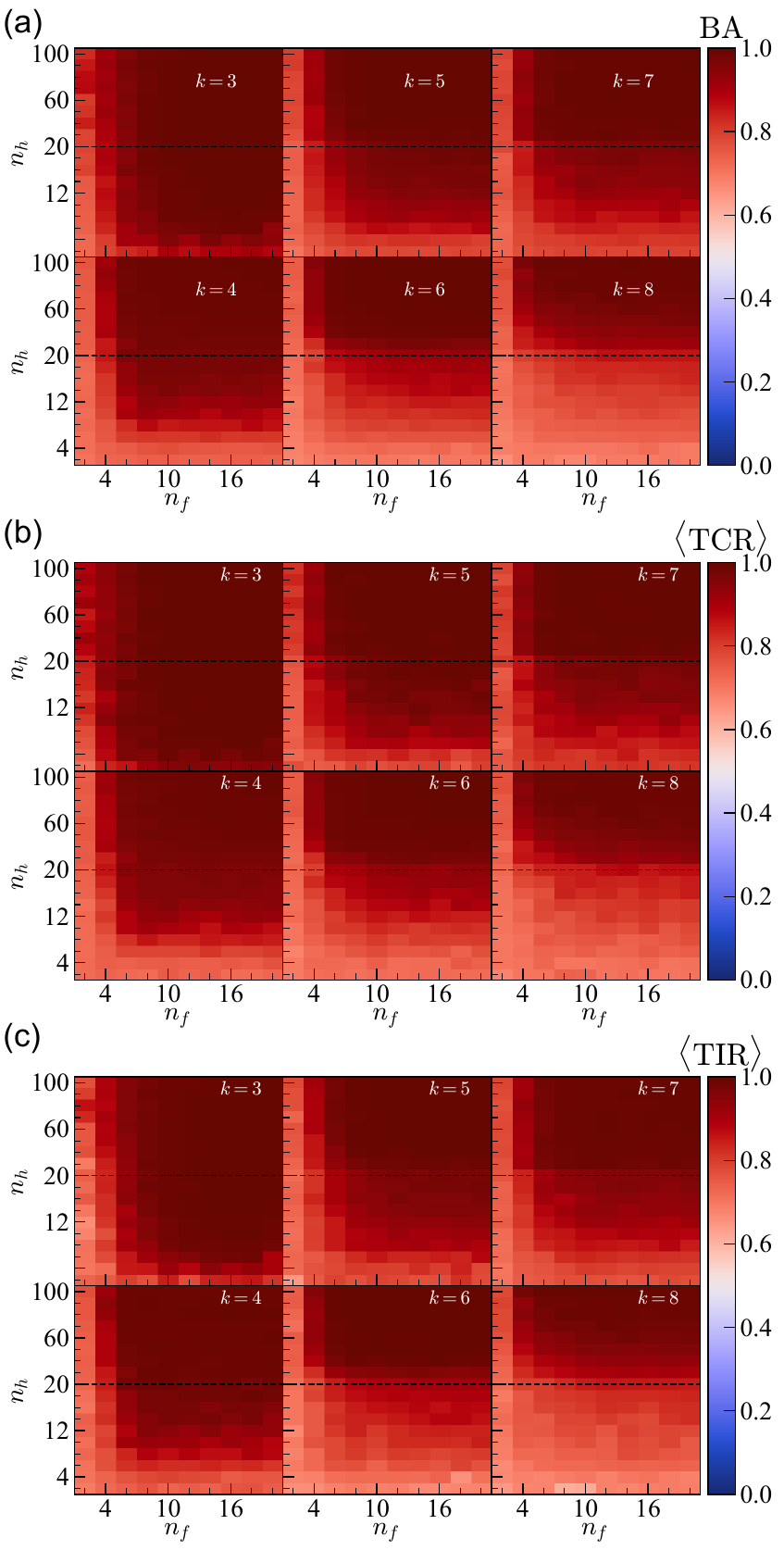}
    \caption{(a) Heatmaps of the fold-averaged balanced accuracy $\mathrm{BA}$ for CNNs with $n_f$ filters and $n_h$ hidden neurons trained on $k\times k$ unit cells indicated on top of each heatmap.
    (b) Heatmaps of the fold-averaged true class C rate $\langle \mathrm{TCR} \rangle$. 
    (c) Heatmaps of the fold-averaged true class I rate $\langle \mathrm{TIR} \rangle$.}
    \label{fig:BA_TPR_TOR}
\end{figure}

Despite the complexity of the classification problem, we find that, for sufficiently large $n_f$ and $n_h$,
the balanced accuracy $\mathrm{BA}$ approaches its maximum value
1 for every considered unit cell size $k$ (Fig.~\ref{fig:BA_TPR_TOR}(a)). Strikingly, the number of filters $n_f$ required to achieve large
BA does not vary with $k$. This is most likely because the plaquettes encode a finite amount of information---there are only 16 unique $2\times 2$ plaquettes. This does not change with unit cell size $k$, thus the required number of filters $n_f$ is invariant to the unit cell size. The number of required hidden neurons $n_h$
increases with $k$, but not dramatically, despite the combinatorial explosion of the design space. To interpret this
result, we note that a high $\mathrm{BA}$ corresponds to correctly
classifying most class C unit cells as class C, and most
class I unit cells as class I. Hence, sufficiently large
networks yield decision boundaries such that most needles are enclosed and most hay is outside (Fig.~1(c, d)). However, whether this decision boundary coarsely (Fig.~1(c)) or finely (Fig.~1(d)) approximates the structure close to the needles cannot be deducted from a coarse measure such as the $\mathrm{BA}$ over the test set.

The usage of $\mathrm{BA}$ to show trends between neural network performance and hyperparameters is warranted, since no significant difference between the true class C rate $\mathrm{TCR}$ and true class I rate $\mathrm{TIR}$ appears to exist, see Fig.~\ref{fig:BA_TPR_TOR}. Evidently $\mathrm{TCR}$ and $\mathrm{TIR}$ depend similarly on the number of filters $n_f$ and number of hidden neurons $n_h$. This is to be expected, since the networks are trained on a class-balanced training set.

\begin{figure}[b]
    \centering
    \includegraphics{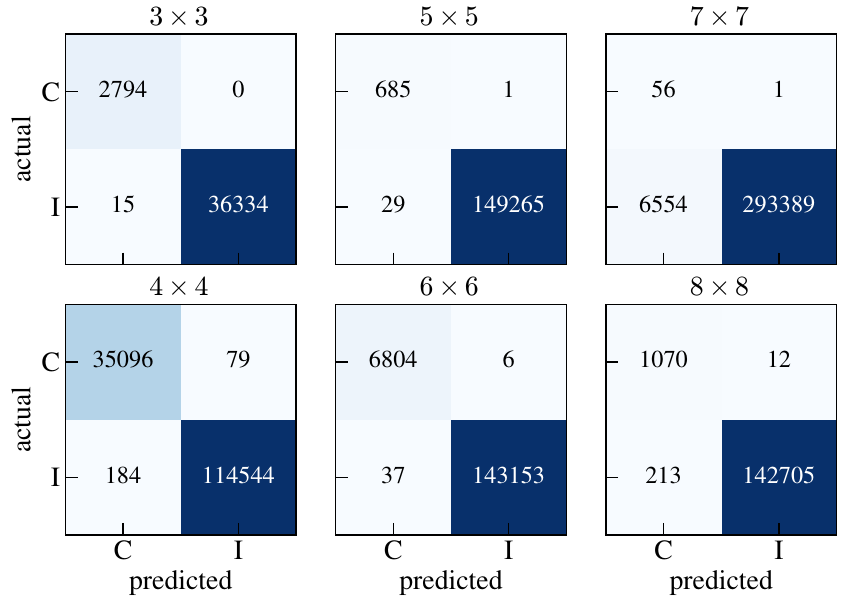}
    \caption{Confusion matrices over the test set for trained CNNs with the highest accuracy over the class-balanced validation set. The $k\times k$ unit cell size is indicated on top of each matrix.}
    \label{fig:F1_CM_testset}
\end{figure}

The effect of class-imbalance on CNN performance can be further illustrated through constructing the confusion matrices (Fig.~\ref{fig:F1_CM_testset}(b)). Though all CNNs show high true C and I rates, the sheer number of falsely classified C unit cells can overtake the number of correctly classified C unit cells if the class-imbalance is sufficiently strong, as for the $7 \times 7$ unit cells. 

\subsubsection{Increasing the size of the training set}
\begin{figure}[b]
    \centering
    \includegraphics{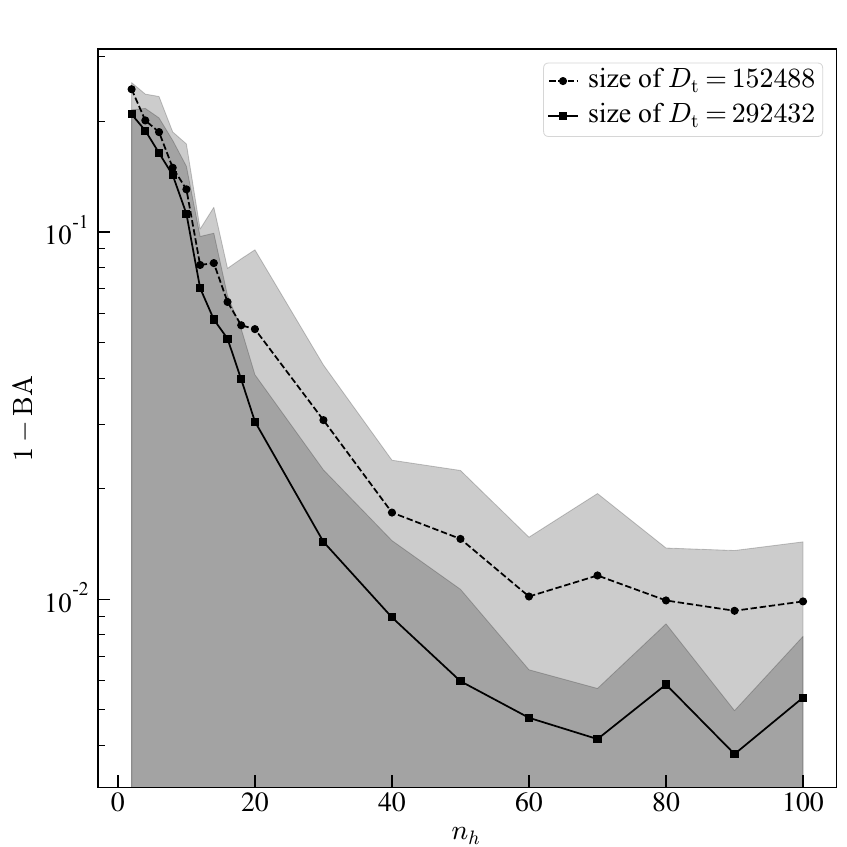}
    \caption{Balanced accuracy $\mathrm{BA}$ for CNNs with $n_f=20$ trained on a smaller training set (circles) and larger training set (squares). The size of the training set is indicated in the legend.}
    \label{fig:ba_7x7_small_big}
\end{figure}

To illustrate how the size of the training set $D_{\mathrm{t}}$ influences the performance over the test set, we compare CNNs trained on two training sets of different size consisting of $7 \times 7$ unit cells---the unit cell size with the strongest class-imbalance. We use the fold-averaged balanced accuracy $\mathrm{BA}$ to quantify the performance. The training sets are obtained from 1M and 2M uniformly sampled unit cells respectively, and the number of class C unit cells is artificially increased using translation and rotation to create class-balanced training sets. The best $\mathrm{BA}$ is more than a factor 2 smaller for CNNs trained on the larger training set, compared to the smaller training set (Fig.~\ref{fig:ba_7x7_small_big}). Thus, lack of performance due to a strong data-imbalance can be improved through increasing the number of training samples.

\subsubsection{Random walk near the class boundary}
To better understand the complexity of the classification problem, we probe the design space near test set unit cell designs. Starting from a test set design $X_0$ with true class C, we rotate a randomly selected unit cell to create a new unit cell design $X_1$. We do this iteratively up to a given number of steps $s$ to create a chain of designs. For each generated design, we assess the new true class using the design rules for classification (i) and through calculating $M(n)$ for $n\in\{3, 4\}$ for classification (ii).

For each unit cell size $k$, we take $s=k^2$ steps in design space. The probability to transition from an initial $5\times 5$ design $X_0$ of class C to another design $X_s$ of class C as a function of $s$ random walk steps in design space $p_{\text C \to \text C}(s)$, is shown in Fig.~3(b, c) for classification problems (i) and (ii). 

\begin{figure}[b!]
    \centering
    \includegraphics{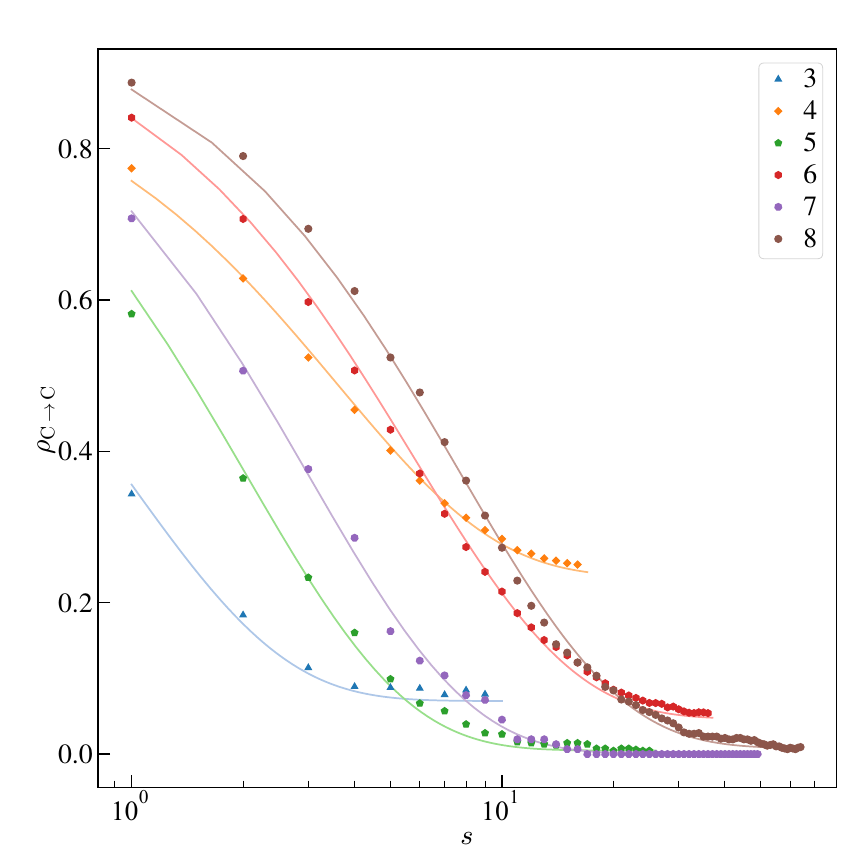}
    \caption{Probability $\rho_{\text C \to \text C}$ (polygons) to transition from initial design $X_0$ of class C to another design $X_s$ of class C as a function of $s$ random walk steps in design space starting from the initial design. The legend indicates the polygon and color for each unit cell size $k$. The continuous lines are obtained from a least-squares fit using Eq.~(1) of the Main Text.}
    \label{fig:p_PtoP_class}
\end{figure}

We repeat the random walks for other $k \times k$ unit cells for problem (i). A clear difference between the different unit cell sizes is visible. Both the rate at which the probability decreases initially, and the value to which it saturates differs per unit cell size (Fig.~\ref{fig:p_PtoP_class}).

For even unit cell size, the dominant strip mode width is $W=1$ (Fig.~\ref{fig:CRS_LM_Schem_PixRep}) and each class C design is most likely to just have a single strip mode. Thus, the probability to transition from C to I relies on the probability to rotate a unit cell inside the strip of the strip mode, which is $1 / k$, so $\alpha_t \approx 1 / k$. For odd unit cell sizes, the dominant strip mode width is $W=2$, such that  $\alpha_t \approx 2 / k$. 

To understand the asymptotic behavior, we note that for large $s$ the unit cells are uncorrelated to their original designs. Thus, the set of unit cells are akin to a uniformly sampled set of unit cells. Consequently, the probability to transition from C to C for large $s$ is approximately equal to the true class C volume fraction $\beta$.

\subsubsection{Random walk near the decision boundary}
In addition to the true class, we can assess the predicted class by a given network for each unit cell in the random walk. This allows us to probe the decision boundary, which is the boundary between unit cells that a given network will classify as C and those it will classify as I. By comparing the transition probabilities for given networks to the true transition probability we get an indication of how close the decision boundary is to the true class boundary.

\begin{figure}[b]
    \centering
    \includegraphics{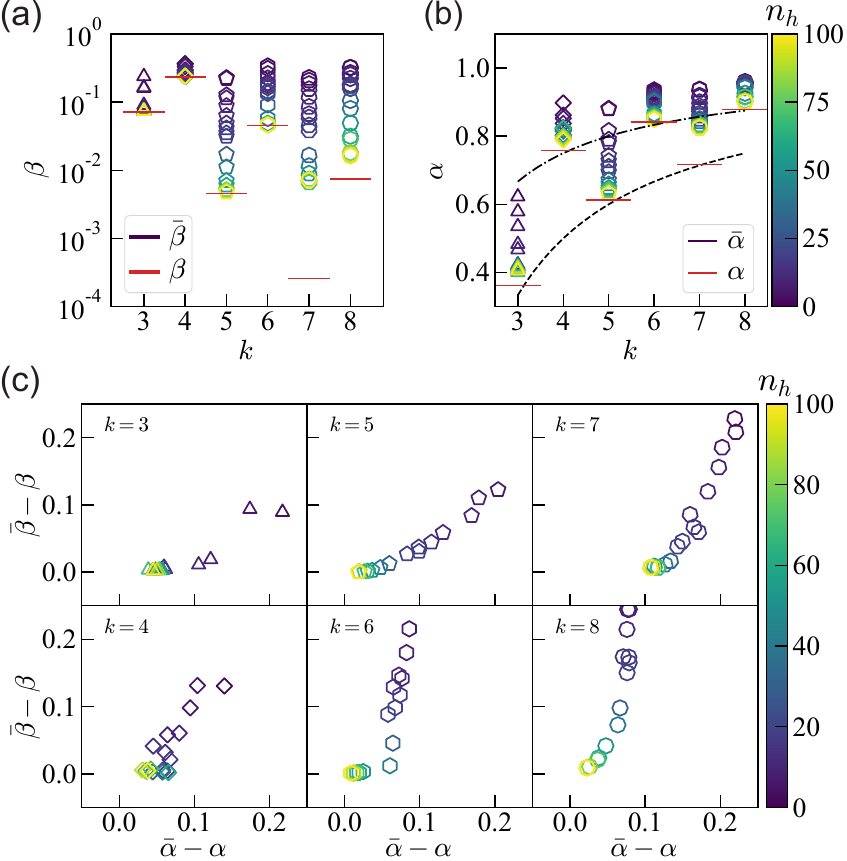}
    \caption{(a) Classification problem (i) Class C volume fraction $\beta$ (red) as
a function of unit cell size $k$. The predicted class C volume fraction $\bar{\beta}(n_h)$ (for $n_f = 20$) approaches $\beta$ for increasing number of hidden neurons $n_h$ (colorbar).
(b) True dimensionality $\alpha$ (red) and predicted dimensionality
$\bar{\alpha} (n_h)$ (colorbar) obtained through least-squares fits to data as in
Fig.~3(b) for all $k$. The estimated $\alpha$ for both odd (dashed
line) and even (dashdotted line) $k$ agree well with $\alpha$.
(c) Scatter plots of class volume fractions
$\bar{\beta}(n_h) - \beta$ versus dimensionality $\bar{\alpha}(n_h) - \alpha$ shows that the
latter asymptotes later than the former ($n_h$ indicated by a
colorbar, and unit cell size $k$ indicated on top of each graph)}
    \label{fig:beta_alpha_k}
\end{figure}

To quantitatively compare the true class boundary with the decision boundaries, we fit the measured transition probability for each network 
to Eq.~(1) of the Main Text with $\bar{\alpha}$ as fitting parameter. We start from designs with true and predicted class C, and track the predicted class for the random walk designs. We set the asymptotic value to the predicted class C volume fraction $\bar{\beta}$ (Fig.~\ref{fig:beta_alpha_k}(a)) for each network. From this we obtain a 10-fold averaged estimate of $\bar{\alpha}$.

Additionally, we do this for varying unit cell size $k$ for classification problem (i) using the hyper parameter grid search networks. We use CNNs with fixed number of filters $n_f=20$ and varying number of hidden neurons $n_h$. We select the networks with the best-performing learning rate $\eta$ over the validation set, and obtain a 10-fold averaged estimate of $\bar{\alpha}$ for each $n_h$ (Fig.~\ref{fig:beta_alpha_k}(b)). 

Small networks tend to overestimate the class C dimensionality $\alpha$ (Fig.~\ref{fig:beta_alpha_k}(b). Larger networks tend to approach the true $\alpha$ for increasing number of hidden neurons $n_h$. For large data-imbalance, as is the case for $k=7$ and $k=8$, even the larger networks overestimate $\alpha$. This is not a fundamental limitation, and can most likely be improved by increasing the size of the training set, see section \emph{Increasing the size of the training set}.
We conjecture that this is due to the higher combinatorial complexity of the C subspace for larger unit cells, which requires a larger number of training samples to adequately learn the relevant features describing the subspace.
The trend shown in Fig.~4(c) holds across all unit cell sizes (Fig.~\ref{fig:beta_alpha_k}(c)).

\section{Computational time analysis}
In this section we discuss the computational time it takes to classify a $k \times k$ unit cell design by calculating the number of zero modes $M(n)$ for $n\in\{2, 3, 4\}$ using rank-revealing QR (rrQR) decomposition. The first algorithm takes as input a unit cell design, creates rigidity matrices $R$ for each $n$, and calculates the dimension of the kernel for each matrix using rrQR decomposition. The classification then follows from the determination of $a$ and $b$ in $M(n) = a n + b$ as described in the main text.

We contrast this brute-force calculation of the class with a trained neural networks time to compute the classification. We consider a shallow CNN with a single convolution layer of $n_f=20$ filters, a single hidden layer of $n_h = 100$ hidden neurons and an output layer of 2 neurons. The network takes as input a $k \times k$ unit cell design in the pixel representation (with padding) and outputs the class. The number of parameters of these CNNs grows with $k$, see eq.\eqref{eq: params CNN}. We focus on networks trained on classification problem (i). 

The brute-force calculation scales nearly cubically with input size $k^2$, while the neural network's computational time remains constant with unit cell size $k$. \ch{This is due to computational overhead---the number of operations for a single forward run of our CNN scales linearly with $k^2$, but can run in parallel on GPU hardware.} This highlights the advantage of using a neural network for classification: it allows for much quicker classification of new designs. In addition, the neural network is able to classify designs in parallel extremely quickly: increasing the number of unit cells to classify from 1 to 1000 only increased the computational time by a factor $\approx 1.33$. 

Please note that this analysis does not include the time to train such neural networks, nor the time it takes to simulate a large enough dataset to train them. Clearly there is a balance, where one has to weigh the time it takes to compute a sufficiently large dataset versus the number of samples that they would like to have classified. For classification problems (i) and (ii) it did not take an unreasonable time to create large enough datasets, yet brute-forcing the entire design space would take too much computational time. Our training sets are large enough to train networks on---of order $10^5$---but are still extremely small in comparison to the total design space, such that the time gained by using a CNN to classify allows for exploring a much larger portion of the design space as generating random designs is computationally cheap.

\begin{figure}[b]
    \centering
    \includegraphics{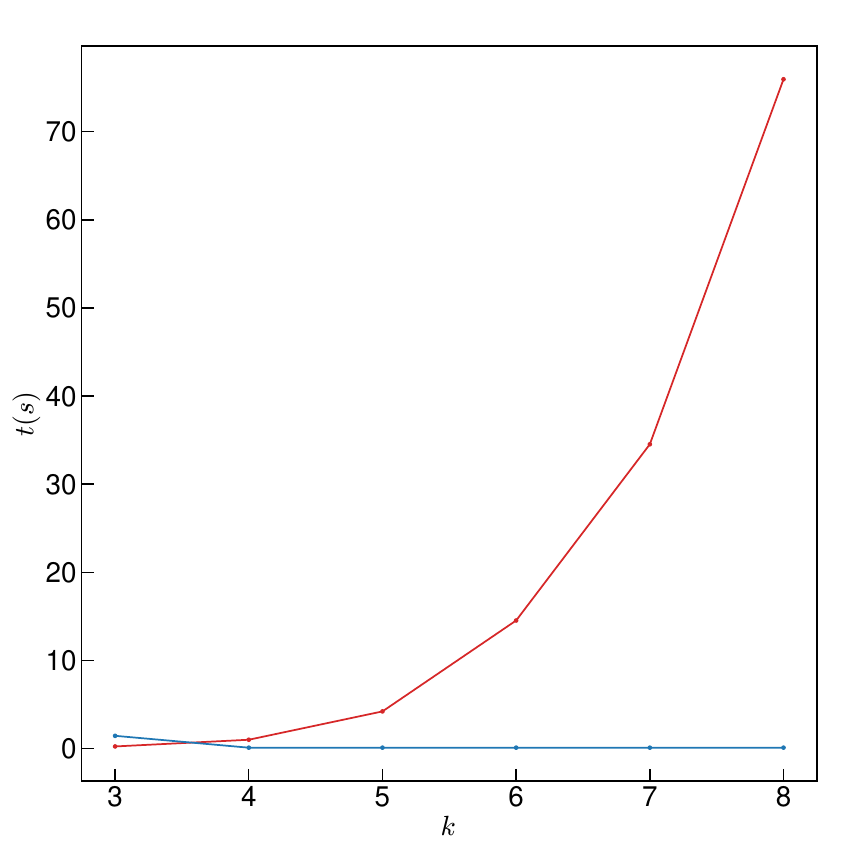}
    \caption{Computation time $t$ measured in seconds $s$ to classify $k \times k$ unit cells by \ch{total number of modes $M(n)$} (red) versus using \ch{a trained convolutional neural network} (blue).}
    \label{fig:time_complexity}
\end{figure}

\end{document}